\documentclass[%
 aip,
rsi,
 amsmath,amssymb,
 reprint,%
]{revtex4-1}

\usepackage{graphicx}
\usepackage{dcolumn}
\usepackage{bm}

\usepackage{upgreek}     
\usepackage{graphics}     
\usepackage{epsf}       
\usepackage{bm}        
\usepackage{changepage}
\usepackage{svg}

\usepackage{hyperref}
\hypersetup{
    colorlinks=true,
    citecolor=magenta,
}

\usepackage[utf8]{inputenc}
\usepackage[T1]{fontenc}
\usepackage{mathptmx}

\begin{document}

\preprint{AIP/123-QED}

\title{Polarization purity for active stabilization of diode laser injection lock}
\author{R. D. Niederriter}
\author{I. Marques Van Der Put}
\author{P. Hamilton}
\email{paul.hamilton@ucla.edu}
\date{\today}
\affiliation{Department of Physics and Astronomy, University of California, Los Angeles, California 90095, USA}

\begin{abstract}
Injection locking of diode lasers is commonly used to amplify low power laser light, but is extremely sensitive to perturbations in the laser current and temperature. To counter such perturbations, active stabilization is often applied to the current of the injection locked diode. We observe that the diode laser's polarization extinction ratio (PER) greatly increases when injection locked, and therefore the PER provides a measure of injection lock quality. We report robust active stabilization of a diode laser injection lock based on the PER, demonstrating the technique at 399~nm wavelength where injection locking is typically less stable than at longer wavelengths. The PER provides a feedback error signal that is compatible with standard PID servo controllers, requires no additional optical components beyond the optical isolator typically used in injection locking, and enables a large feedback bandwidth. 
\end{abstract}

\maketitle

\section{\label{sec:Intro}Introduction}

Injection locking of laser diodes is commonly used for generating the high optical power and narrow frequency spectrum required for many experiments in atomic, molecular, and optical (AMO) physics \cite{InjectionLocking_BlueDiodes_2003}. 
Injection locking uses a low power primary (seed) laser coupled into the laser cavity of a high power secondary laser diode. 
With no seed power, the high power diode lases on multiple longitudinal modes with typical spacing on the order of 100 GHz, leading to a broad spectrum. 
If the narrow-linewidth seed light is near a resonance of the secondary laser's cavity, mode competition suppresses gain at other frequencies and enables narrow-linewidth high-power output. 

The challenge for laser diode injection locking is stability, which requires maintaining a close match between the secondary laser diode's resonance and the primary laser diode's frequency. 
In AMO applications, the seed laser frequency is typically fixed (for example, referenced to an atomic transition), and the secondary laser diode's current and temperature are varied to achieve injection locking. 

Some injection locked systems achieve passive stability, typically using high seed power \cite{InjectionLocking_Passive_399nm_2015}, but in many cases active stabilization is needed. 
Active stabilization relies on measuring the quality of the injection lock and adjusting the secondary diode laser parameters to maintain optimal injection locking. 
Injection locking can narrow the secondary laser's optical frequency spectrum \cite{ScanningCavity_Gupta_2016}, reduce the intensity noise and frequency noise \cite{TrackedResonances_2007}, change the total output power \cite{PowerModulationLocking_2008}, and, as investigated in this work, change the polarization purity. 
Each of these parameters contains information about the quality of the injection lock and can be used in a feedback loop to maximize the quality of the lock. 
For example, the optical frequency spectrum can be measured with a spectrometer such as a scanning Fabry-Perot cavity \cite{ScanningCavity_Gupta_2016} or a grating spectrometer \cite{GratingFilteredInjectedLaser_2015}, the intensity noise can be measured using a photodiode and RF power detector \cite{TrackedResonances_2007}, and the polarization purity can be measured using a polarizing beam splitter. 

In this work, we propose and demonstrate a technique for actively stabilizing injection locking based on the polarization extinction ratio (PER). The PER can be measured quickly (typically limited by photodetector bandwidth) and can produce a zero-crossing error signal for simple analog feedback.
Furthermore, measuring the PER of an injection locked laser is almost free, as typical injection locking setups use polarizing beam splitters as part of optical isolators. The power reflected from one of these polarizing beam splitters is inversely proportional to the PER and is a convenient proxy for injection lock quality, meaning no additional optical components are necessary. 

\begin{figure}[b]
    \centering
    \includegraphics[width=\columnwidth]{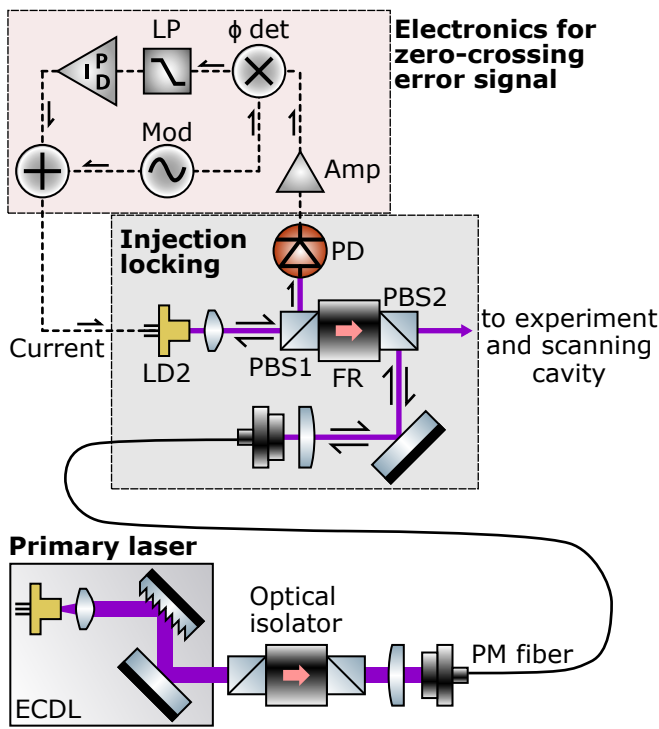}
    \caption{Experimental schematic.
    The primary laser (ECDL) output passes through an optical isolator into a polarization-maintaining fiber to seed the secondary laser diode (LD2). 
    The seed light is directed backwards through the optical isolator (PBS1, FR, and PBS2).
    The secondary laser's polarization purity is measured using the reflection from PBS1, detected by a photodiode (PD).
    Half-wave-plates (not shown) are used to maximize transmission of the secondary laser output through the isolator and to direct the seed light into LD2. 
    A scanning confocal cavity (not shown) is used to measure the single-frequency content of the secondary laser spectrum for diagnostics only and is not used to stabilize the injection lock performance. 
    To produce a zero-crossing error signal, a small current modulation (Mod) is applied to the secondary laser diode (LD2); the modulated polarization purity is detected using the AC-coupled output of PD, an amplifier (Amp), a phase detector ($\phi$~det), and a low-pass filter (LP). 
    A home-built PI controller completes the feedback loop to maintain optimal injection locking.
    } 
    \label{fig:SetUp}
\end{figure}

In contrast, using a spectrometer introduces bulky optical components, has slow measurement rate ($\sim$100~Hz), and may not be cost-effective when multiple injection locked lasers are needed. 
Measuring the noise doesn't indicate which direction to change the current \cite{TrackedResonances_2007}, so a slow scan and lock-in amplifier are typically needed.
The secondary laser's output power changes quickly and modulation produces an error signal that can be used with fast analog servo controllers \cite{PowerModulationLocking_2008}. However, the signal is small: the laser power typically changes by about 1~mW on a baseline of about 100~mW. 
Meanwhile, the PER changes quickly by a factor on the order of 10, resulting in a much larger signal size; the PER is also amenable to modulation and demodulation techniques to generate a zero-crossing error signal. 
The PER feedback bandwidth is in the kHz to MHz range (limited by electronics such as the photodetector bandwidth, PID controller, etc.), allowing orders of magnitude faster response than typically possible with a spectrometer.

Others have noted that the polarization of light output by a laser diode can be controlled or purified using feedback from an external cavity \cite{ECDL_Polarization_TE&TMswitching_1994} or when injecting an external optical signal \cite{LD_PolarizationControl_InjectionLocking_1977}. 
We extend these ideas to use the PER for active stabilization of the injection lock. 

We note that any injection-lock quality measurement could be used with a microcontroller to stabilize the injection lock \cite{ScanningCavity_Gupta_2016} (in contrast to an analog servo controller). Using the PER with a microcontroller would still allow much faster feedback, as the measurement rate is much higher than is achievable using a spectrometer.

\section{Apparatus}

Our injection locking system illustrated in Fig.~\ref{fig:SetUp} is very similar to those used by others \cite{ScanningCavity_Gupta_2016,GratingFilteredInjectedLaser_2015,InjectionLocking_BlueDiodes_2003}, with the addition of a single photodetector and feedback electronics.
The primary laser is an external cavity diode laser (ECDL) with a laser diode (Nichia NDV4B16-E, free running wavelength of 400~nm) and a 3600~lines/mm grating separated by $\approx$3~cm.
Light from the ECDL is coupled through a polarization maintaining (PM) fiber to seed the secondary laser (also Nichia NDV4B16-E). 
The seed power used is 5~mW for a secondary laser output power of $\approx$150~mW. 

The secondary laser output passess through an optical isolator comprised of two polarizing beam splitters (PBS1 and PBS2) and a Faraday rotator (FR).
A half-wave-plate before the isolator aligns the secondary laser polarization to maximize transmission through the isolator. Power reflected from PBS1 is sent to a photodiode (PD) to measure the PER. A small amount of power from the secondary laser is reflected by PBS2. This reflected beam is coupled into the same PM fiber as the seed laser; the narrow alignment tolerance of fiber coupling ensures the seed light is nearly optimally coupled into the secondary laser (LD2). 
A half-wave-plate (not shown) aligns the seed polarization to pass backwards through the optical isolator. 
The seed laser spatial mode profile is also shaped by a cylindrical lens telescope (not shown) to match the secondary laser for better injection locking. A small portion of the secondary laser output is sent into a scanning confocal Fabry-Perot cavity to monitor the injection locking performance. 
We scan the length of the confocal cavity using a piezo actuator and measure the cavity transmission with a photodiode; the maximum photodiode signal is proportional to the frequency purity of the secondary laser.
The scanning cavity is diagnostic only and is not used to stabilize the injection lock as in previous work \cite{ScanningCavity_Gupta_2016}.

The new technique we describe here relies on measuring the PER of the secondary laser diode to quantify the injection lock quality.
We measure the polarization purity using the optical isolator's first PBS; the PBS has PER of $\approx$3000 (35~dB), well above that of the laser diode. The PER is defined as the ratio of the transmitted to reflected power ($\rm{P}_{\rm{trans}}$ and $\rm{P}_{\rm{refl}}$, respectively), 
\begin{equation}
    \rm{PER} = \frac{P_{\rm{trans}}}{P_{\rm{refl}}}.
    \label{eq:PER}
\end{equation}
Over a small current range, $\rm{P}_{\rm{trans}}$ is nearly constant and therefore $\rm{P}_{\rm{refl}} \propto 1/\rm{PER}$.
While the PER is the more fundamental quantity of interest, it requires measuring both $\rm{P}_{\rm{trans}}$ and $\rm{P}_{\rm{refl}}$. In practice, feedback is based on $\rm{P}_{\rm{refl}}$ measured by a single photodetector (PD in Fig~\ref{fig:SetUp}). 

\begin{figure}[t]
    \centering
    \includegraphics[width=\columnwidth]{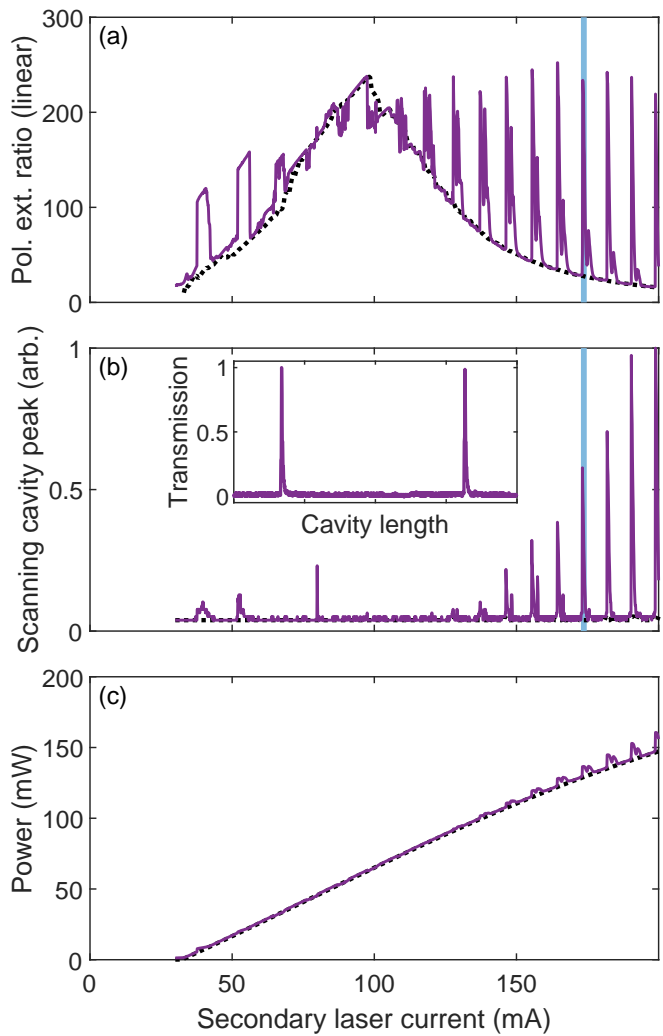}
    \caption{(a) Measured polarization purity (polarization extinction ratio, PER), 
    (b) scanning cavity peak height, 
    and (c) secondary laser output power vs injected laser current with (solid lines) and without (dotted black lines) seed power. 
    Inset: Scanning cavity transmission showing a full free spectral range. The scanning cavity peak height in (b) is extracted from the scanning cavity transmission.
    Injection locking occurs in narrow intervals spaced by $\approx$9~mA.
    The free-running PER increases at moderate power and then decreases at high power. Injection locking increases the PER by almost an order of magnitude, and the laser power slightly increases as well. 
    The shaded region corresponds to the current range of Fig.~\ref{fig:PER_ZoomedIn}.
    The seed power was 5~mW. 
    } 
    \label{fig:PER_FullScale}
\end{figure}

The measured PER of the secondary laser is shown in Fig.~\ref{fig:PER_FullScale}, along with the secondary laser output power (measured after the isolator).
The PER of the free-running diode is low near the laser threshold ($\approx$35~mA), increases initially, and then decreases at higher currents, as shown by the dashed-line. 
The observed low PER at high current is consistent with previous reports of high power laser diode polarization purity \cite{LD_PolarizationPurity_SPIE_2018}.

\begin{figure}[t!]
    \centering
    \includegraphics[width=\columnwidth]{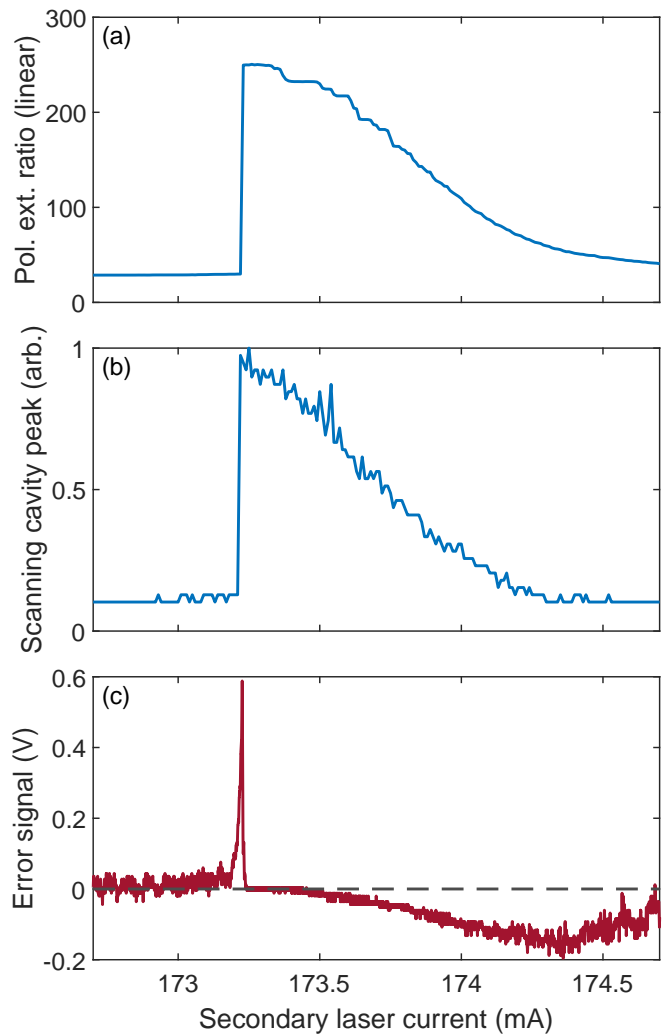}
    \caption{Zoomed-in version of Fig.~\ref{fig:PER_FullScale}. (a) Measured polarization purity (polarization extinction ratio, PER), (b) scanning cavity peak height, and (c) demodulated error signal vs injected laser current with (solid red lines) and without (dotted black lines) seed power. 
    Injection locking increases the scanning cavity peak height (with a similar shape reported by others \cite{ScanningCavity_Gupta_2016}), also greatly increases the PER, and slightly increases the laser power. 
    The demodulated error signal has a zero crossing near the peak and enables analog feedback to stabilize the secondary laser current for optimum injection locking (ex: PI or PID).
    The seed power was 5~mW and the secondary laser modulation current was 28~$\upmu$A$_{\rm{pp}}$ at 100~kHz.
    } 
    \label{fig:PER_ZoomedIn}
\end{figure}

Injection locking greatly increases the PER within a narrow current range, such as that shown in Fig.~\ref{fig:PER_ZoomedIn}. 
The peaks in the PER correspond to the same current regions where the scanning cavity peak height increases (proportional to the power in a single optical frequency), indicating that an increase in PER is a good metric of injection lock quality.
In addition, we observe the secondary laser output power increases slightly when injection locked. 
The observed asymmetry in the diode laser injection locking curve is consistent with theory\cite{InjectionLockingSemiconductorLaser_Theory_Lang_1982,InjectionLockingSemiconductorLaser_Theory_Li_1994b} and previous results
\cite{ScanningCavity_Gupta_2016}. 
To generate a zero-crossing error signal from the measured PER, we add a small modulation to the secondary laser current and then demodulate $\rm{P}_{\rm{refl}}$ using a phase detector.
Typical modulation parameters are 28~$\upmu$A$_{\rm{pp}}$ at 100~kHz (pp: peak-to-peak).
The resulting error signal is shown in Fig.~\ref{fig:PER_ZoomedIn}(c). 
Ideally, the zero-crossing (for example, at $\approx$173.2~mA in Fig.~\ref{fig:PER_ZoomedIn}) corresponds to the maximum PER and therefore the best injection locking. In practice, the demodulation electronics add a small DC offset, changing the zero-crossing location. To compensate, we introduce an adjustable DC offset (5--20~mV) which allows us to control the precise location of the zero-crossing to optimize the injection locking performance. 

The error signal is compatible with standard feedback schemes including PID controllers. 
The PER or demodulated error signal could also be used as the input to digital feedback schemes or maximization algorithms implemented on an FPGA or microcontroller \cite{ScanningCavity_Gupta_2016}.

The small current modulation used for generating a zero-crossing error signal leads to a small modulation in the laser power. Using a modulation current of 28~$\upmu$A$_{\rm{pp}}$  results in a power variation less than 100~$\upmu$W$_{\rm{pp}}$. 
This power modulation is negligible for many applications. 
If power modulation needs to be avoided in a particular application, a maximization algorithm without modulation can be used instead. 
Note that the secondary laser frequency is fixed by the seed frequency, and we do not observe frequency modulation associated with the applied current modulation.

\section{Results}

\begin{figure}
    \centering
    \includegraphics[width=\columnwidth]{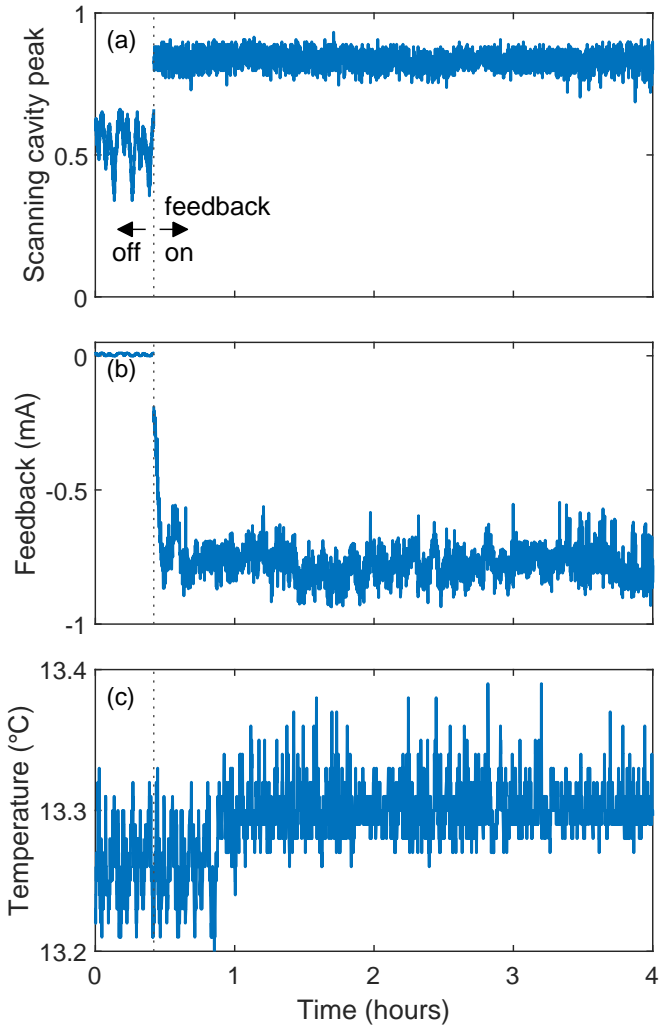}
    \caption{Injection locking performance for several hours with and without feedback. 
    (a) Scanning cavity peak height; (b) feedback from the PI controller to stabilize the injection lock; (c) secondary laser temperature. The PI feedback was initially disabled, then feedback was enabled at 25~minutes (marked by dotted vertical line). Without feedback, the injection locking parameters (such as scanning cavity peak height and polarization extinction ratio) fluctuate. Feedback stabilizes these parameters. The secondary laser current and output power were approximately 195~mA and 150~mW, respectively.
    The seed power was 5~mW and the secondary laser modulation current was 28~$\upmu$A$_{\rm{pp}}$ at 100~kHz.
    } 
    \label{fig:vsTime}
\end{figure}

\begin{figure}
    \centering
    \includegraphics[width=\columnwidth]{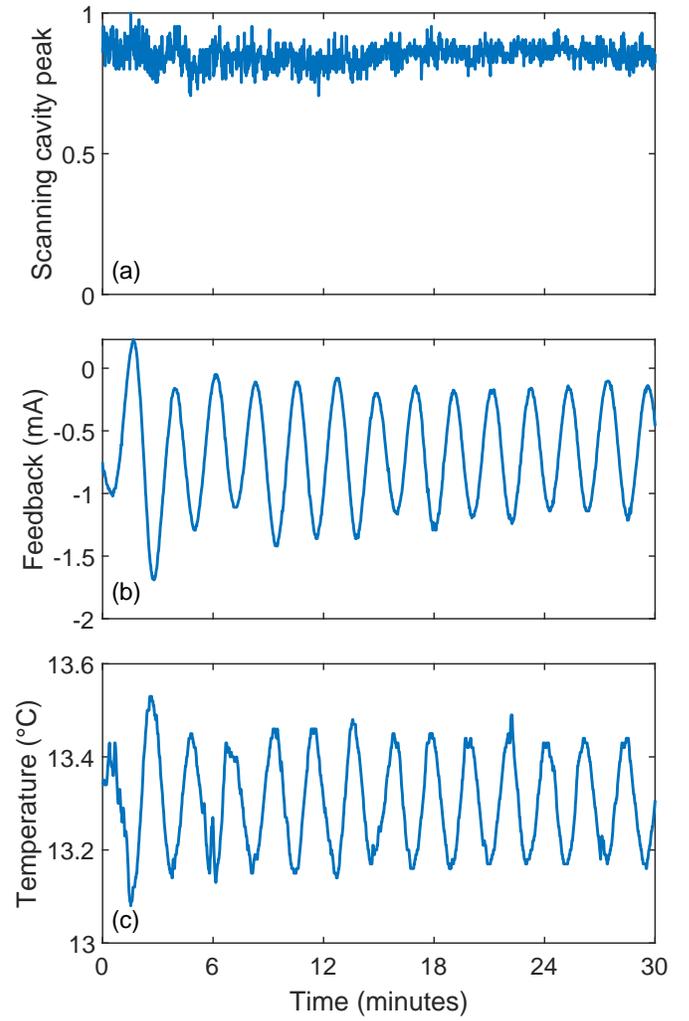}
    \caption{Injection locking performance with feedback while the laser temperature was oscillating. (a) Scanning cavity peak height; (b) feedback from the PI controller to stabilize the injection lock; (c) secondary laser temperature. Feedback based on the demodulated PER stabilizes the scanning cavity peak height. The secondary laser current and output power were approximately 173~mA and 130~mW, respectively.
    The seed power was 5~mW and the secondary laser modulation current was 28~$\upmu$A$_{\rm{pp}}$ at 100~kHz.
    } 
    \label{fig:Oscillating_vsTime}
\end{figure}

The stabilized injection lock performance is shown in Fig.~\ref{fig:vsTime}. 
For reference, the same parameters were recorded without feedback before enabling the PI feedback loop. 
Without feedback, the secondary laser output power at the seed frequency showed large fluctuations. In addition, the injection lock was often lost if the current was set too close to the peak of the injection locking curve (for example, near 173.2~mA in Fig.~\ref{fig:PER_ZoomedIn}); to prevent losing the lock altogether, the current needed to be set about 0.5~mA above the optimum injection locking point. 

With PI feedback based on the demodulated PER, the injection locking performance was much more stable  compared to free-running without feedback. 
Despite temperature fluctuations of more than 100~mK, the PI feedback held the injection-locked laser frequency purity constant over several hours (characterized by the scanning cavity peak height and PER). 
The demonstrated performance is sufficient for AMO physics applications; we use this polarization-stabilized injection locked laser system for a Zeeman slower on the 399~nm $^1$S$_0$ to $^1$P$_1$ transition in Yb \cite{CavityProbe_2020}. 
We typically use the system all day without adjustment of the injection lock. Without feedback, however, we typically need to adjust the secondary laser current every 20 minutes.

To further demonstrate the ability of the stabilization technique, we deliberately caused large oscillations in the secondary laser's temperature; results are shown in Fig.~\ref{fig:Oscillating_vsTime}. 
The secondary laser's temperature is controlled by a commercial TEC controller. Increasing the temperature controller's PID settings caused the temperature to oscillate with 2.1~minute period and peak-to-peak amplitude of $\approx$0.3~K. 
Accordingly, the feedback system adjusted the secondary laser diode's current to compensate. The large temperature change required $>$1~mA change in current. 
Comparing to Fig.~\ref{fig:PER_ZoomedIn}, such a large change in current corresponds to a very large change in the injection lock frequency purity and would typically cause the injection lock to fail.
The feedback scheme maintains the injection lock even with large changes in temperature. 

\section{Conclusion}

We have shown that the PER of an injection locked laser diode is a useful metric for the quality of the injection lock. 
Using an error signal derived from the PER, we implemented a simple and robust technique to actively tune the laser's current in order to maintain the injection lock for long periods of time. 
The zero-crossing error signal enables using either analog or digital feedback, including standard PID controllers. 
As this technique requires no additional optical components, it introduces no loss of laser power,  requires little space in optical setups, and is inexpensive, making it well suited to systems with multiple injection locked lasers.

\begin{acknowledgments}
The authors acknowledge the generous support of this work by ONR Grant N000141712407 and DARPA Grant D18AP00067.
\end{acknowledgments}

\section*{Data Availability}
The data that support the findings of this study are available from the corresponding author upon reasonable request.

\bibliography{InjectionLocking} 

\end{document}